\documentclass[aps,prl,twocolumn,superscriptaddress]{revtex4-1}
\usepackage{graphicx}
\usepackage{amssymb}
\usepackage{amsmath}
\usepackage{hyperref}
\usepackage{color}

\newcommand{\ket}[1]{|#1 \rangle}

\newcommand{\pd}{\partial}

\newcommand{\lr}[1]{\langle #1\rangle}
\newcommand{\diag}{\mathrm{diag}}

\newcommand{\mr}[1]{\mathrm{#1}}
\newcommand{\mbb}[1]{\mathbb{#1}}

\newcounter{fnnumber}

\begin{document}
\title{Conformal Boundary Conditions of Symmetry-Enriched Quantum Critical Spin Chains}

\author{Xue-Jia Yu}
\altaffiliation{These authors contributed equally.}
\affiliation{International Center for Quantum Materials, School of Physics, Peking University, Beijing 100871, China}
\author{Rui-Zhen Huang}
\altaffiliation{These authors contributed equally.}
\affiliation{Kavli Institute for Theoretical Sciences and CAS Center for Excellence in Topological Quantum Computation, University of Chinese Academy of Sciences, Beijing, 100190, China}
\author{Hong-Hao Song}
\affiliation{Kavli Institute for Theoretical Sciences and CAS Center for Excellence in Topological Quantum Computation, University of Chinese Academy of Sciences, Beijing, 100190, China}
\author{Limei Xu}
\affiliation{International Center for Quantum Materials, School of Physics, Peking University, Beijing 100871, China}
\affiliation{Collaborative Innovation Center of Quantum Matter, Beijing, China}
\affiliation{Interdisciplinary Institute of Light-Element Quantum Materials and Research Center for Light-Element Advanced Materials, Peking University, Beijing, China}
\author{Chengxiang Ding}
\affiliation{School of Science and Engineering of Mathematics and Physics, Anhui University of Technology, Maanshan, Anhui 243002, China}
\author{Long Zhang}
\email{longzhang@ucas.ac.cn}
\affiliation{Kavli Institute for Theoretical Sciences and CAS Center for Excellence in Topological Quantum Computation, University of Chinese Academy of Sciences, Beijing, 100190, China}

\date{\today}

\begin{abstract}
Some quantum critical states cannot be smoothly deformed into each other without either crossing some multicritical points or explicitly breaking certain symmetries even if they belong to the same universality class. This brings up the notion of ``symmetry-enriched'' quantum criticality. While recent works in the literature focused on critical states with robust degenerate edge modes, we propose that the conformal boundary condition (b.c.) is a more generic characteristic of such quantum critical states. We show that in two families of quantum spin chains, which generalize the Ising and the three-state Potts models, the quantum critical point between a symmetry-protected topological phase and a symmetry-breaking order realizes a conformal b.c. distinct from the simple Ising and Potts chains. Furthermore, we argue that the conformal b.c. can be derived from the bulk effective field theory, which realizes a novel bulk-boundary correspondence in symmetry-enriched quantum critical states.
\end{abstract}
\maketitle

\emph{Introduction.}---The development of topological states of matter has greatly deepened our understanding of gapped phases \cite{Wen2017}. For example, one-dimensional (1D) symmetry-protected topological (SPT) states are fully classified by the projective representations of the symmetry group and host degenerate edge modes, which transform as the projective representations \cite{Gu2009, Pollmann2010}. Different SPT phases cannot be adiabatically connected without either crossing phase transitions or explicitly breaking the symmetry.

Quantum critical states enjoy scale invariance in the low-energy limit, and fall into different universality classes characterized by the operator scaling dimensions. Surprisingly, some quantum critical states cannot be smoothly connected by tuning model parameters without either crossing some multicritical points or explicitly breaking certain symmetries even if they belong to the same universality class. This brings up the notion of gapless SPT \cite{Keselman2015, Scaffidi2017a} or ``symmetry-enriched'' quantum critical states \cite{Verresen2019}. Robust degenerate edge states persist in some quantum critical states \cite{Cheng2011, Fidkowski2011, Sau2011, Kestner2011, Iemini2015, Lang2015, Keselman2015, Ruhman2017, Scaffidi2017a, Jiang2018, Parker2018, Keselman2018, Verresen2018, Verresen2019, Jones2019, Verresen2020, Thorngren2021}, which are secured by the symmetry-flux (disorder) operators in the bulk carrying nontrivial symmetry charges \cite{Verresen2019}. This indicates a novel bulk-boundary correspondence. However, a signature of symmetry-enriched quantum critical states without edge degeneracy is still lacking.

In this work, we shall show that the conformal boundary condition (b.c.) and the associated surface critical behavior are more generic characteristics of symmetry-enriched quantum criticality. For a critical system with boundary, a conformal b.c. corresponds to a fixed point of the renormalization group (RG) flow of the boundary states. Different conformal b.c. can be specified for a given conformal field theory (CFT) in the bulk, resulting into rich surface critical phenomena.

The conformal b.c. determines the operator content of the system \cite{Cardy1986a, Cardy1989}, i.e., the Hamiltonian eigenstates, which are organized into conformal families, each comprising a primary state and all its conformal descendants. The conformal b.c. also determines the universality of the surface criticality \cite{Cardy1984, Cardy1986a}. Given a local operator on the boundary [denoted by $\phi(r)$] and in the bulk [$\phi_{b}(R)$], the connected correlation functions scale as
\begin{align}
&C_{\parallel}(r_{1}-r_{2}) = \langle\phi(r_{1})\phi(r_{2})\rangle_{c} \propto |r_{1}-r_{2}|^{-2\Delta_{\phi}}, \label{eq:Cpara} \\
&C_{\perp}(r-R) = \langle\phi(r)\phi_{b}(R)\rangle_{c} \propto |r-R|^{-\Delta_{\phi}-\Delta_{\phi}^{b}}, \label{eq:Cperp}
\end{align}
in which $\langle AB\rangle_{c}=\langle AB\rangle-\langle A\rangle\langle B\rangle$. Here, $r_{1}-r_{2}$ is parallel to the surface, while $r-R$ is perpendicular to it. $\Delta_{\phi}$ and $\Delta_{\phi}^{b}$ are the scaling dimensions of the boundary and the bulk operators, respectively. The surface critical behavior of classical statistical systems has been extensively studied \cite{Binder1983phase, Diehl1986phase}. The interest in the surface criticality has been revived recently motivated by the impact of topological edge states at quantum critical points (QCPs), leading to the discovery of new universality classes \cite{Grover2012a, Suzuki2012b, Zhang2017, Ding2018, Weber2018, Weber2019a, Weber2021, Zhu2021, Ding2021, ParisenToldin2021, Hu2021, ParisenToldin2022, Jian2021, Metlitski2022, Padayasi2022}.

In this work, we study two families of quantum spin chains, which generalize the 1D Ising and the three-state Potts models \cite{Verresen2018, Jones2019, OBrien2020}, respectively. Each family contains quantum critical states that are described by the same CFT but cannot be smoothly connected. While the two Ising QCPs can be distinguished by the degenerate edge states \cite{Verresen2018, Jones2019, Verresen2019}, the generalized Potts chain lacks such a distinctive feature. By examining their surface critical behavior and the energy and entanglement spectra, we show that in each family of models, the QCP between an SPT and a symmetry-breaking order realizes a conformal b.c. distinct from the simple Ising and Potts models. Moreover, the conformal b.c. can be derived from the effective field theory of the bulk states, thus establishing a novel bulk-boundary correspondence in symmetry-enriched quantum critical states.

\emph{Warm-up: Quantum Ising chains.}---We first study the following transverse-field Ising (TFI) chain and the cluster Ising (CI) chain \cite{Jones2019},
\begin{align}
H_{\mathrm{TFI}} &= -\sum_{l=1}^{L-1}\sigma_{l}^{z}\sigma_{l+1}^{z}+h\sum_{l=1}^{L}\sigma_{l}^{x}, \\
H_{\mathrm{CI}} &= -\sum_{l=1}^{L-1}\sigma_{l}^{z}\sigma_{l+1}^{z}-h\sum_{l=1}^{L-2}\sigma_{l}^{z}\sigma_{l+1}^{x}\sigma_{l+2}^{z}.
\end{align}
Both models enjoy the $\mathbb{Z}_{2}\times\mathbb{Z}_{2}^{T}$ symmetry generated by $\Pi_{x}=\prod_{l=1}^{L}\sigma_{l}^{x}$ and $T=\mathcal{K}$ (the complex conjugation). There is a QCP at $h_{c}=1$ in both models separating the ferromagnetic (FM) order at $|h|<h_{c}$ and the disordered phase at $h>h_{c}$. The QCPs are described by the 2D Ising CFT.

\begin{figure}[tb]
\includegraphics[width=\columnwidth]{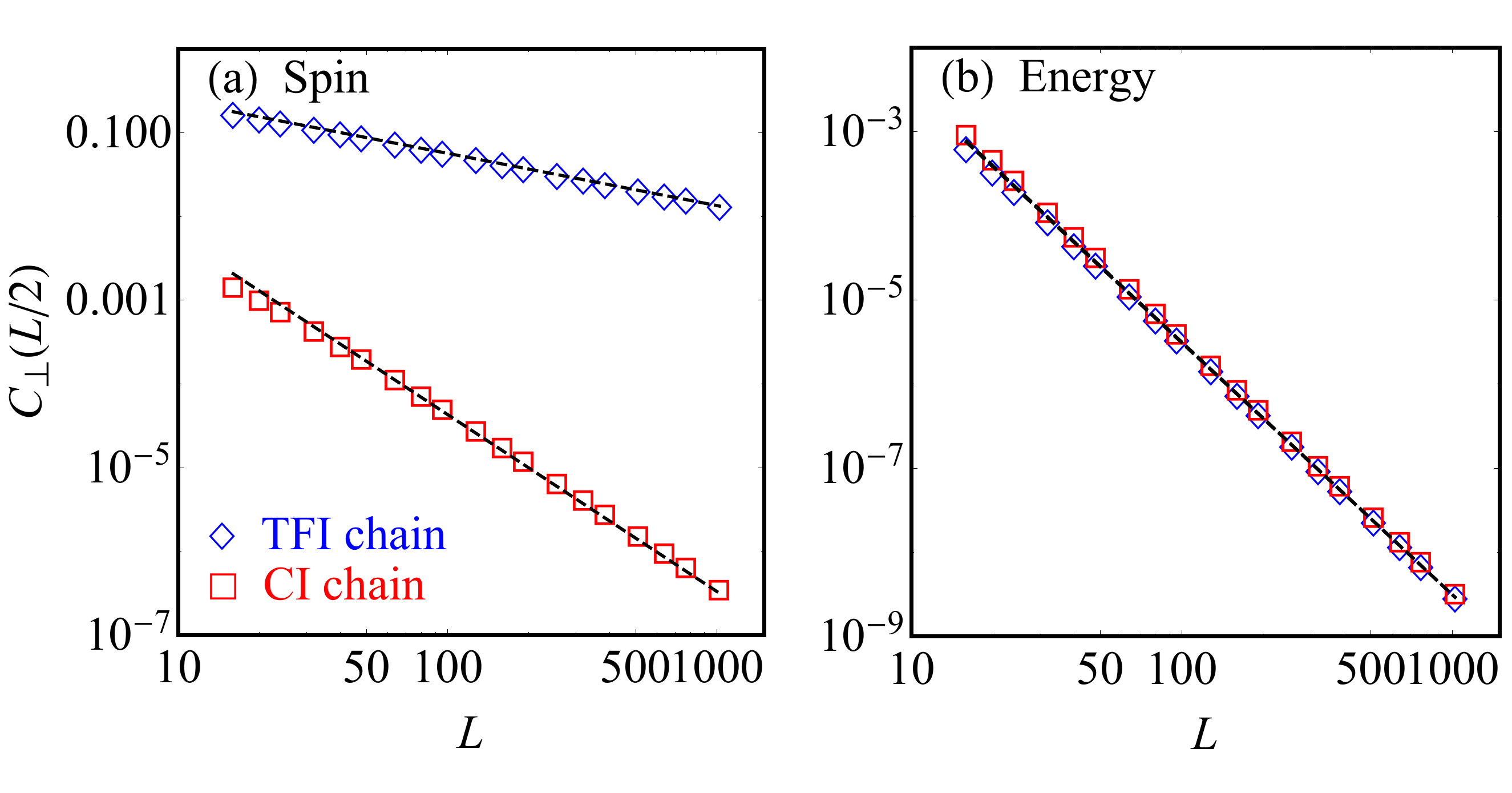}
\caption{Connected correlation functions $C_{\perp}(L/2)$ of (a) the spin operator $\sigma_{l}^{z}$ and (b) the energy operator $\epsilon_{l}=\sigma_{l}^{z}\sigma_{l+1}^{z}$ of the critical Ising chains. Dashed lines are the power-law fitting according to Eq. (\ref{eq:Cperp}).}
\label{fig:Ising}
\end{figure}

Both models are exactly solvable with the Jordan-Wigner transformation \cite{Jones2019}, $\sigma_{l}^{z}=\prod_{k=1}^{l-1}(i\gamma_{k}\tilde{\gamma}_{k})\gamma_{l}$ and $\sigma_{l}^{x}=i\gamma_{l}\tilde{\gamma}_{l}$, in which the Majorana fermion operators satisfy $\{\gamma_{k},\gamma_{l}\}=\{\tilde{\gamma}_{k},\tilde{\gamma}_{l}\}=2\delta_{kl}$, and $\{\gamma_{k},\tilde{\gamma}_{l}\}=0$. In the Majorana representation,
\begin{equation}
H=-\sum_{l=1}^{L-1}i\tilde{\gamma}_{l}\gamma_{l+1}-h\sum_{l=1}^{L-\alpha}i\tilde{\gamma}_{l}\gamma_{l+\alpha},
\end{equation}
in which $\alpha=0$ for the TFI and $2$ for the CI chain. At the QCP, both models are mapped to 1D massless Majorana fermions. However, in the CI chain, there are two decoupled Majorana modes denoted by $\gamma_{1}$ and $\tilde{\gamma}_{L}$, which lead to a two-fold degeneracy in each energy level \cite{Verresen2018, Jones2019}. These edge zero modes are protected by the $\mathbb{Z}_{2}^{T}$ symmetry \cite{Verresen2019}. In the spin representation, the degeneracy comes from the conservation of the edge spin operators $\sigma_{1}^{z}=\gamma_{1}$ and $\sigma_{L}^{z}=-i\tilde{\gamma}_{L}\Pi_{x}$, which label the edge magnetization. The energy spectrum falls into four sectors labeled by $\sigma_{1}^{z}$ and $\sigma_{L}^{z}$. Therefore, the $\mathbb{Z}_{2}$ symmetry is spontaneously broken on the edges.

The connected correlation functions $C_{\perp}(L/2)$ of the spin operator $\sigma_{l}^{z}$ and the energy operator $\epsilon_{l}=\sigma_{l}^{z}\sigma_{l+1}^{z}$ are used to characterize the surface critical behavior. These correlations are calculated in the Majorana representation, which is detailed in the Supplemental Materials \footnote{See the Supplemental Materials, which also include Refs. \cite{Verstraete2008, Verstraete2006, Haegeman2017, Zauner-Stauber2018, Rader2018, Corboz2018, Calabrese2004, Calabrese2009b, Alba2018, Kitaev2006, Levin2006}.}\setcounter{fnnumber}{\thefootnote}, and fitted to Eq. (\ref{eq:Cperp}) with the bulk scaling dimensions $\Delta_{\sigma}^{b}=1/8$ and $\Delta_{\epsilon}^{b}=1$ (see Fig. \ref{fig:Ising}). The extracted scaling dimensions of the boundary operators are listed in Table \ref{tab:exponents}. While the TFI chain is captured by the Ising CFT with free b.c., the exponents of the CI chain are consistent with the fixed b.c., which can be attributed to the spontaneous edge magnetization. This feature should also apply to other 1D quantum critical states with edge degeneracy.

\begin{ruledtabular}
\begin{table}[tb]
\caption{Scaling dimensions of the boundary spin and energy operators. Scaling dimensions in the Ising and Potts boundary CFTs are also listed for comparison.}
\label{tab:exponents}
\begin{tabular}{cc|cc}
Class	& Model/b.c.			& $\Delta_{\sigma}$	& $\Delta_{\epsilon}$	\\
\hline
Ising	& TFI					& 0.4992(3)			& 1.99957(8)			\\
		& CI					& 1.9984(2)			& 2.00008(2)			\\
CFT		& Free					& 1/2				& 2						\\
		& Fixed					& 2					& 2						\\
\hline
Potts	& Disorder-FM			& 0.66598(3)		& 0.7993(2)				\\
		& Disorder-NotA	    	& 0.6629(1)			& 0.7915(5)				\\
		& RSPT-FM				& 0.0661(3)			& 0.204(9)				\\
		& RSPT-NotA				& 0.0601(1)			& 0.21(1)				\\
CFT		& Free					& 2/3				& 4/5					\\
		& Dual-mixed			& 1/15				& 1/5
\end{tabular}
\end{table}
\end{ruledtabular}

\begin{figure}[tb]
\includegraphics[width=\columnwidth]{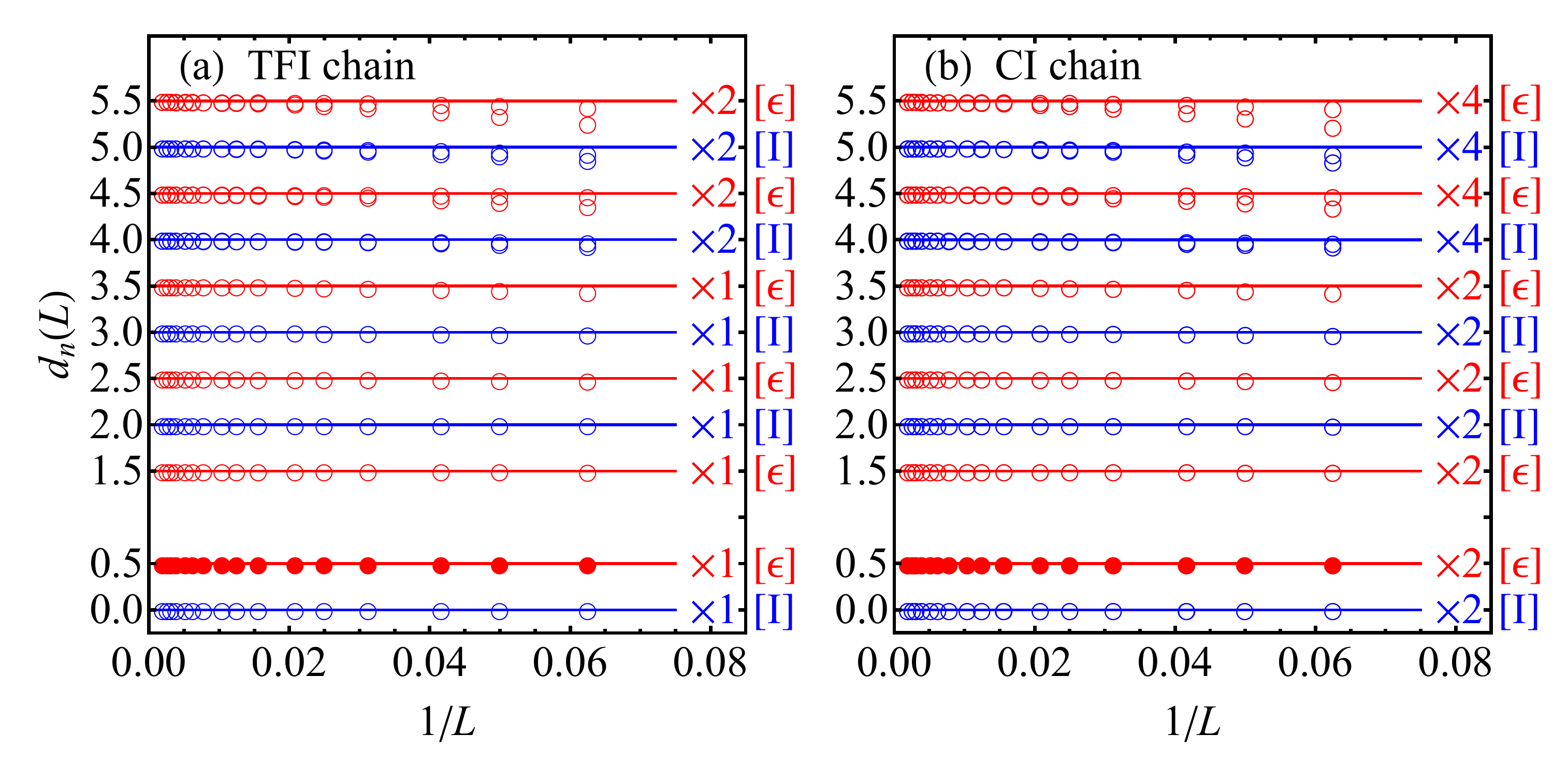}
\caption{Low-energy spectra of (a) the TFI and (b) the CI chains with open boundaries. The excitation energy $\Delta \epsilon_{n}(L)=\epsilon_{n}(L)-\epsilon_{0}(L)$ is normalized to the effective scaling dimension, $d_{n}(L)=\frac{1}{2}\frac{\Delta\epsilon_{n}(L)}{\Delta\epsilon_{1}(L)}$, such that the first excited state (marked with filled circles) maps to $1/2$. The conformal family and the expected degeneracy in the boundary CFT are labeled on the right. The operator content in (a) is $[\mbb{I}]\oplus[\epsilon]$, consistent with the Ising CFT with free b.c. \cite{Cardy1986a}. In (b), the energy levels in blue come from $(\sigma_{1}^{z},\sigma_{L}^{z})=(+,+)$ and $(-,-)$ sectors, each forming a conformal family $[\mbb{I}]$, while those in red from $(+,-)$ and $(-,+)$ sectors, each forming $[\epsilon]$. This is consistent with the fixed b.c. \cite{Cardy1986a}.}
\label{fig:IsingSpec}
\end{figure}

The energy spectra with open boundaries are shown in Fig. \ref{fig:IsingSpec}. With proper normalization, the excitation energies are mapped to the operator scaling dimensions and compared with the operator content of the boundary CFT, which is explained in the figure caption. These spectra are fully consistent with the Ising CFT with free b.c. and fixed b.c., respectively.

\emph{Generalized Potts chain: Recapitulation.}---We then study the 1D generalized three-state Potts model introduced in Ref. \cite{OBrien2020}. Following their notation, the Hamiltonian is given by
\begin{equation}
H=H_{\mr{P}}+\lambda H_{0}, \label{eq:Potts}
\end{equation}
in which $H_{\mr{P}}$ is the simple quantum Potts model,
\begin{equation}
H_{\mr{P}}=-J\sum_{l=1}^{L-1}(\sigma_{l}^{\dag}\sigma_{l+1}+\sigma_{l}\sigma_{l+1}^{\dag})-f\sum_{l=1}^{L}(\tau_{l}+\tau_{l}^{\dag}),
\end{equation}
and $H_{0}$ is given by
\begin{equation}
H_{0}=\sum_{l=1}^{L-1}3\big((S_{l}^{+}S_{l+1}^{-})^{2}-S_{l}^{+}S_{l+1}^{-}+\mathrm{H.c.}\big)-\sum_{l=1}^{L}(\tau_{l}+\tau_{l}^{\dag}).
\end{equation}
The operators are defined by $\tau=\mr{diag}(1,\omega,\omega^{2})$ with $\omega=e^{2\pi i/3}$, and
\begin{equation}
\sigma=
\begin{pmatrix}
0 & 1 & 0 \\
0 & 0 & 1 \\
1 & 0 & 0
\end{pmatrix},\quad
S^{+}=
\begin{pmatrix}
0 & 0 & 1 \\
1 & 0 & 0 \\
0 & 0 & 0
\end{pmatrix}=(S^{-})^{\dagger}.
\end{equation}
$S^{\pm}$ are the ladder operators of $S^{z}=\diag(0,1,-1)$. The model (\ref{eq:Potts}) enjoys the $S_{3}$ symmetry generated by the $\mathbb{Z}_{3}$ rotation $R=\prod_{l=1}^{L}\tau_{l}$ and the charge conjugation $C=\prod_{l=1}^{L}c_{l}$, in which
\begin{equation}
c=
\begin{pmatrix}
1 & 0 & 0 \\
0 & 0 & 1 \\
0 & 1 & 0
\end{pmatrix}.
\end{equation}

\begin{figure}[tb]
\includegraphics[width=\columnwidth]{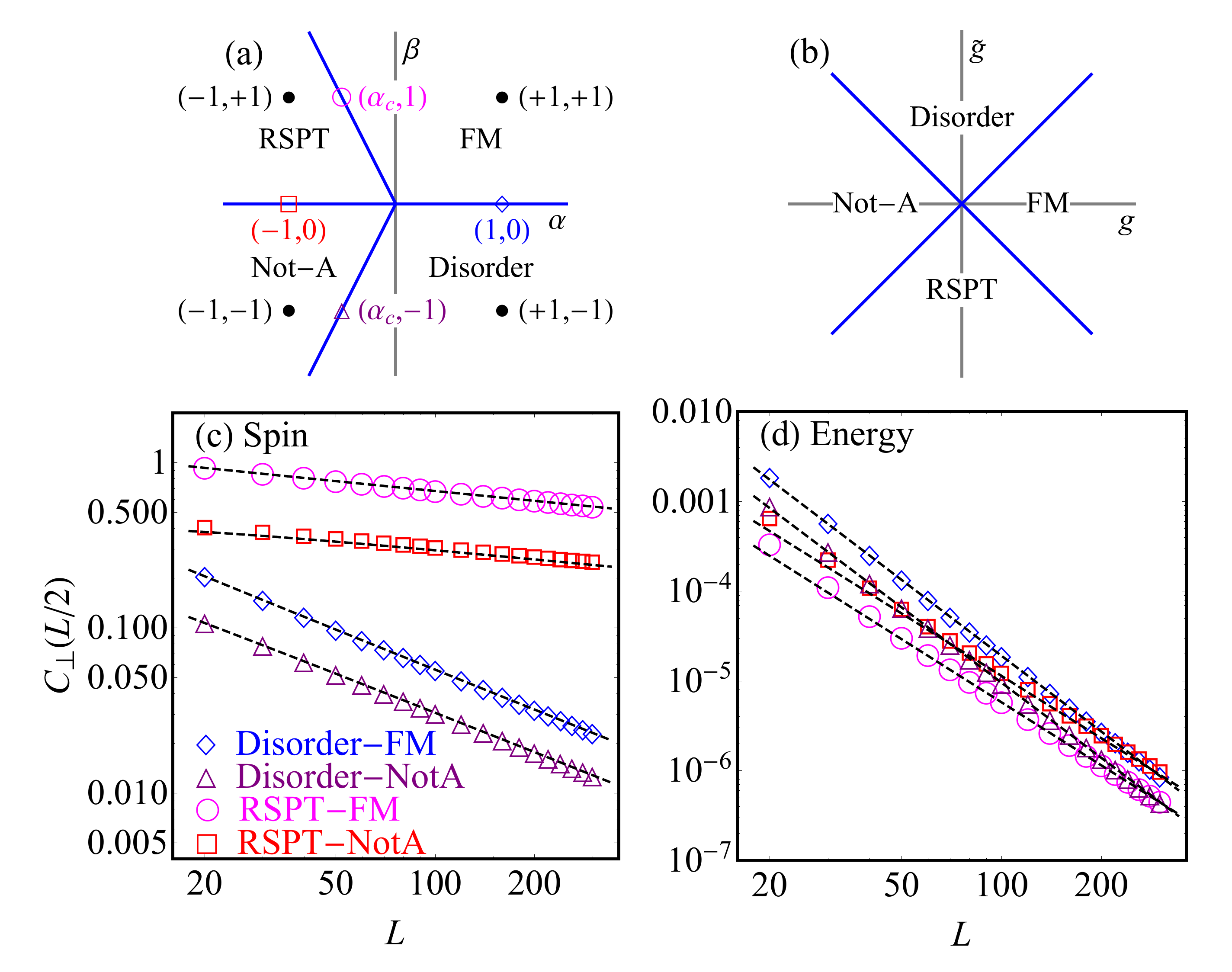}
\caption{(a) Schematic quantum phase diagram of the generalized Potts chain. The special points $(\alpha,\beta)=(\pm 1,\pm 1)$ are marked by filled circles. The QCPs studied in this work are indicated on the transition lines, in which $\alpha_{c}=-0.50509$. (b) Global phase diagram of the effective field theory (\ref{eq:field}). (c,d) Connected correlation functions $C_{\perp}(L/2)$ of the spin operator $\sigma_{l}$ and the energy operator $\epsilon_{l}=\tau_{l}+\tau_{l}^{\dag}$ at the QCPs. Dashed lines are the power-law fitting according to Eq. (\ref{eq:Cperp}) plus a correction-to-scaling term, $bL^{-\Delta_{\phi}-\Delta_{\phi}^{b}-1}$.}
\label{fig:Potts}
\end{figure}

\begin{figure*}[tb]
\includegraphics[width=\textwidth]{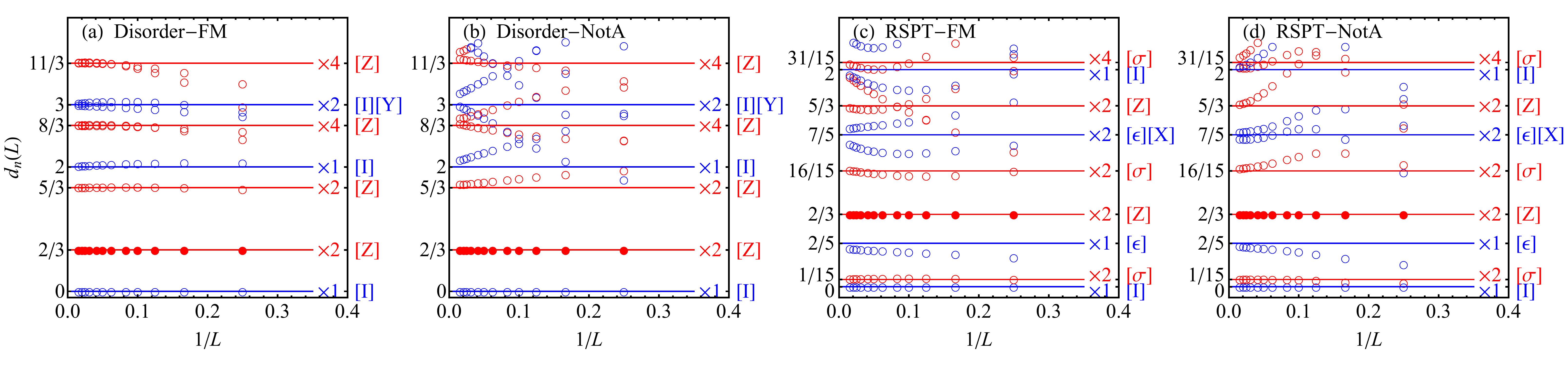}
\caption{Low-energy spectra of the generalized Potts chain at the QCPs. The excitation energy $\Delta\epsilon_{n}(L)=\epsilon_{n}(L)-\epsilon_{0}(L)$ is normalized to the effective scaling dimension, $d_{n}(L)=\frac{2}{3}\frac{\Delta\epsilon_{n}(L)}{\Delta\epsilon_{Z}(L)}$, such that the primary state in $[Z]$ (marked with filled circles) is normalized to $2/3$. The conformal family and the expected degeneracy in the boundary CFT are labeled on the right. Red circles indicate numerically exactly two-fold degenerate levels, while blue circles indicate non-degenerate levels. The disorder-FM and the disorder-NotA QCPs are consistent with the operator content of the Potts CFT with free b.c. \cite{Cardy1986a}, $[\mathbb{I}]\oplus [Y]\oplus 2[Z]$, while the RSPT-FM and the RSPT-NotA QCPs are consistent with that of the dual-mixed b.c. \cite{Affleck1998, Fuchs1998}, $[\mathbb{I}]\oplus [Y]\oplus[\epsilon]\oplus [X]\oplus 2[\sigma]\oplus 2[Z]$.}
\label{fig:PottsSpecRev}
\end{figure*}

Let us first recapitulate the phase diagram of the model (\ref{eq:Potts}), and refer to Ref. \cite{OBrien2020} for the detailed derivation. The quantum phase diagram parametrized by $\lambda=1-\alpha$, $J=\alpha+\beta$, and $f=\alpha-\beta$ is sketched in Fig. \ref{fig:Potts} (a), which shows four gapped phases. Each phase is adiabatically connected to a special point within the phase, $(\alpha,\beta)=(\pm 1,\pm 1)$, where the ground states can be exactly constructed. The Potts ferromagnetic (FM) ordered phase has three-fold degenerate ground states, which are smoothly connected to the fully polarized FM states $\otimes_{l}\ket{A}_{l}$, $\otimes_{l}\ket{B}_{l}$ and $\otimes_{l}\ket{C}_{l}$, in which $\ket{A}$, $\ket{B}$ and $\ket{C}$ are the eigenstates of $\sigma$ with eigenvalues $1$, $\omega$ and $\omega^{2}$, respectively. The FM order is characterized by the order parameter $\lr{\sigma_{l}}^{3}>0$. The disordered phase is smoothly connected to the $S_{3}$-symmetric state $\otimes_{l}\ket{0}_{l}$, in which $\ket{0}=\frac{1}{\sqrt{3}}(\ket{A}+\ket{B}+\ket{C})$ is an eigenstate of $S^{z}$. The other two phases are quite unconventional. The ``not-$A$'' ordered phase contains $(\alpha,\beta)=(-1,-1)$, where the three-fold degenerate ground states are given by $\otimes_{l}\ket{\bar{A}}_{l}$, $\otimes_{l}\ket{\bar{B}}_{l}$ and $\otimes_{l}\ket{\bar{C}}_{l}$, with $\ket{\bar{A}}=\frac{1}{\sqrt{2}}(\ket{B}+\ket{C})$, $\ket{\bar{B}}=\frac{1}{\sqrt{2}}(\ket{C}+\ket{A})$ and $\ket{\bar{C}}=\frac{1}{\sqrt{2}}(\ket{A}+\ket{B})$. The $S_{3}$ symmetry is spontaneously broken with the order parameter $\lr{\sigma_{l}}^{3}<0$. The last phase is dubbed the representation SPT (RSPT) state. At $(\alpha,\beta)=(-1,1)$, its ground state can be constructed with the matrix-product state $R=\otimes_{l}R_{l}$, in which
\begin{equation}
R_{l}=
\begin{pmatrix}
\ket{0}_{l} & \ket{+}_{l}	\\
\ket{-}_{l}	& \ket{0}_{l}
\end{pmatrix}.
\end{equation}
Here, $\ket{0}$, $\ket{+}$ and $\ket{-}$ are the three eigenstates of $S^{z}$. The ground state is given by $\mr{tr}(R)$ for periodic b.c., which is $S_{3}$-symmetric and nondegenerate. With open b.c., the matrix elements of $R$ are four-fold degenerate ground states, which form a linear representation of the $S_{3}$ symmetry.

The duality transformation $\tau_{l}\mapsto \sigma_{l}^{\dagger}\sigma_{l+1}$, $\sigma_{l}\mapsto \prod_{j=1}^{l}\tau_{j}$ exchanges the two terms in $H_{\mr{P}}$ and leaves $H_{0}$ invariant, thus maps the FM to the disordered phase, and the not-$A$ order to the RSPT phase, and vice versa. There is a continuous quantum phase transition between each ordered phase and each disordered phase, all of which correspond to the spontaneous $S_{3}$-symmetry breaking and belong to the 2D three-state Potts universality class. These transition lines join at the multicritical point $(\alpha,\beta)=(0,0)$, which is self-dual and has the U(1) symmetry generated by $Q=\sum_{l}S^{z}_{l}$.

The phase diagram is captured by the following self-dual sine-Gordon theory with the Hamiltonian density [see Fig. \ref{fig:Potts} (b)],
\begin{equation}
\mathcal{H}=\frac{3}{4\pi}(\pd_{x}\phi)^{2}+\frac{3}{4\pi}(\pd_{x}\theta)^{2}-g\cos(3\phi)-\tilde{g}\cos(3\theta), \label{eq:field}
\end{equation}
in which $\phi$ and $\theta$ are the scalar field and the dual disorder field, respectively. The multicritical point is captured by the free boson theory with a compactification radius $\sqrt{3/2}$, which has the same U(1) symmetry and self-duality \cite{Baranowski1990, OBrien2020}. The $g$ and $\tilde{g}$ terms are relevant at the multicritical point. For $g>|\tilde{g}|>0$, $\phi$ is polarized to $0$ or $\pm 2\pi/3$ and gives the FM order, while for $g<-|\tilde{g}|<0$, $\phi$ is polarized to $\pi$ or $\pm \pi/3$, i.e., an equal-weight superposition of two out of the three spin states, and describes the not-$A$ order. For $\tilde{g}>|g|>0$, the disorder field $\theta$ is pinned at $0$ or $\pm 2\pi/3$, while for $\tilde{g}<-|g|<0$, $\theta$ is pinned at $\pi$ or $\pm \pi/3$. From the duality relation, the former corresponds to the disordered phase, while the latter is the RSPT phase. The transition lines are given by $|g|=|\tilde{g}|$, which cannot be smoothly connected without either crossing the multicritical point or explicitly breaking the $S_{3}$ symmetry despite that they belong to the same universality class. However, unlike the CI chain, there are not any degenerate edge states at these QCPs, thus we must find new characteristics.

\emph{Generalized Potts chain: Conformal b.c.}---We first study the surface critical behavior. The connected correlation functions of the spin operator $\sigma_{l}$ and the energy operator $\epsilon_{l}=\tau_{l}+\tau_{l}^{\dag}$ are calculated with the density-matrix renormalization group (DMRG) algorithm \cite{White1992, Schollwock2005, Schollwock2011} detailed in \footnotemark[\thefnnumber] and shown in Fig. \ref{fig:Potts}. The scaling dimensions of the boundary operators are extracted by fitting Eq. (\ref{eq:Cperp}) with $\Delta_{\sigma}^{b}=2/15$ and $\Delta_{\epsilon}^{b}=2$ and listed in Table \ref{tab:exponents}. The four QCPs fall into two classes. While the disorder-FM and the disorder-NotA transitions are captured by the Potts CFT with free b.c. \cite{Cardy1986a}, the RSPT-FM and the RSPT-NotA transitions yield different critical exponents. It turns out that these exponents can be derived from the Potts CFT with the ``new'' conformal b.c. discovered in Refs. \cite{Affleck1998, Fuchs1998}. This b.c. is $S_{3}$ symmetric and dual to the mixed b.c. \footnote{With the mixed b.c., the edge spin is polarized as an equal-weight superposition of two out of the three spin states with the edge order \unexpanded{$\langle\sigma\rangle^{3}<0$}.}, thus we call it the dual-mixed b.c.

In order to support the identification of these conformal b.c., we show the low-energy spectra at these QCPs with open b.c. in Fig. \ref{fig:PottsSpecRev}, which are normalized to the effective scaling dimensions. The spectra of the disorder-FM and the disorder-NotA QCPs are consistent with the operator content of the Potts CFT with free b.c. On the other hand, the spectra of the RSPT-FM and the RSPT-NotA QCPs are consistent with the dual-mixed b.c.

The operator content of a boundary CFT also shows up in the entanglement spectrum of the ground state, which is equivalent to the energy spectrum of the CFT with a proper conformal b.c. specified at the entangling surface, i.e., the boundary between the subsystem and the rest of the chain \cite{Lauchli2013, Ohmori2015, Cardy2016}. The entanglement spectra of the generalized Potts chain are shown in the Supplemental Materials \footnotemark[\thefnnumber], which points to the same operator contents as the corresponding energy spectra.

These distinct conformal b.c. can be derived from the bulk effective field theory (\ref{eq:field}). Given that the disorder-FM transition at $g=\tilde{g}>0$ realizes the free b.c. \cite{Cardy1986a}, the dual field $\theta$ is pinned at $0$ or $\pm 2\pi/3$ at the boundary due to the duality between the free b.c. and the fixed b.c. \cite{Druhl1982}. At the RSPT-FM transition for $\tilde{g}=-g<0$, the sign of $\tilde{g}$ is reversed, thus $\theta$ is pinned at $\pi$ or $\pm \pi/3$ at the boundary, corresponding to the mixed b.c. of the $\theta$ field and thus the dual-mixed b.c. in terms of the $\phi$ field. On the other hand, the conformal b.c. is not changed by reversing the sign of $g$, because the $\phi$ field is not polarized at the boundary for any of these critical states. Therefore, the disorder-NotA transition at $g=-\tilde{g}<0$ realizes the free b.c., while the RSPT-NotA transition at $g=\tilde{g}<0$ shows the dual-mixed b.c. These arguments are consistent with our numerical results, thereby we establish a novel bulk-boundary correspondence of the symmetry-enriched quantum critical states.

\emph{Discussions.}---The critical exponents of the boundary correlation functions can be measured with the surface magnetic susceptibility etc. \cite{Binder1983phase}. Besides, the characteristic energy spectrum manifests itself in thermodynamic quantities. The thermal entropy of a critical chain of length $L$ at temperature $T$ scales as \cite{Affleck1991}
\begin{equation}
S(L,T)=\frac{\pi c}{3v}LT+\ln g,
\end{equation}
in which $c$ is the central charge, $v$ is the velocity in the low-energy limit, and $\ln g$ is the Affleck-Ludwig boundary entropy, which is a universal constant determined by the conformal b.c. The excess boundary entropy of a conformal b.c. compared with the fixed b.c. can be extracted from the entropy released upon applying a magnetic field on the boundary. If there are degenerate edge states, they contribute an integer factor of degeneracy to $g$, while $g$ is not an integer for a generic conformal b.c. but may be taken as an ``effective edge degeneracy''. Therefore, the conformal b.c. generalizes the edge degeneracy in characterizing quantum critical states.

\emph{Conclusion.}---To summarize, we have studied the generalized Ising and Potts chains. In each family, we focused on the quantum critical states that cannot be smoothly connected even though they are captured by the same CFT. We showed that a distinct conformal b.c. is realized at the QCPs of SPT states. The conformal b.c. is a more generic characteristic of symmetry-enriched quantum critical states beyond the degenerate edge states.

\begin{acknowledgements}
We thank Paul Fendley, Shang Liu, Rong-Yang Sun, Huajia Wang, and Yijian Zou for helpful discussions and communications. Part of the numerical simulations was carried out with the ITensor package \cite{Fishman2020}. This work is supported by the National Key R\&D Program of China (2018YFA0305800), the National Natural Science Foundation of China (11935002, 11975024, 12047554, 12174387, 11774002, and 11804337), the Strategic Priority Research Program of CAS (XDB28000000), China Postdoctoral Science Foundation (2020T130643), Anhui Provincial Supporting Program for Excellent Young Talents in Colleges and Universities (gxyqZD2019023), the Fundamental Research Funds for the Central Universities, and the CAS Youth Innovation Promotion Association.
\end{acknowledgements}

\bibliography{../../BibTex/library,../../BibTex/books}
\end{document}


\title{Conformal Boundary Conditions of Symmetry-Enriched Quantum Critical Spin Chains: Supplemental Materials}

\author{Xue-Jia Yu}
\altaffiliation{These authors contributed equally.}
\affiliation{International Center for Quantum Materials, School of Physics, Peking University, Beijing 100871, China}
\author{Rui-Zhen Huang}
\altaffiliation{These authors contributed equally.}
\affiliation{Kavli Institute for Theoretical Sciences and CAS Center for Excellence in Topological Quantum Computation, University of Chinese Academy of Sciences, Beijing, 100190, China}
\author{Hong-Hao Song}
\affiliation{Kavli Institute for Theoretical Sciences and CAS Center for Excellence in Topological Quantum Computation, University of Chinese Academy of Sciences, Beijing, 100190, China}
\author{Limei Xu}
\affiliation{International Center for Quantum Materials, School of Physics, Peking University, Beijing 100871, China}
\affiliation{Collaborative Innovation Center of Quantum Matter, Beijing, China}
\affiliation{Interdisciplinary Institute of Light-Element Quantum Materials and Research Center for Light-Element Advanced Materials, Peking University, Beijing, China}
\author{Chengxiang Ding}
\affiliation{School of Science and Engineering of Mathematics and Physics, Anhui University of Technology, Maanshan, Anhui 243002, China}
\author{Long Zhang}
\email{longzhang@ucas.ac.cn}
\affiliation{Kavli Institute for Theoretical Sciences and CAS Center for Excellence in Topological Quantum Computation, University of Chinese Academy of Sciences, Beijing, 100190, China}

\begin{abstract}
In these Supplemental Materials, we show the calculation details on the correlation functions of the generalized Ising chains, and the quantum critical point, the correlation functions and the entanglement spectra of the generalized three-state Potts chain.
\end{abstract}

\date{\today}

\maketitle

\section{Correlation functions of the generalized Ising chains}

The TFI and the CI chains can be mapped to free Majorana fermions with the Jordan-Wigner transformation \cite{Jones2019}, $\sigma_{l}^{z}=\prod_{k=1}^{l-1}(i\gamma_{k}\tilde{\gamma}_{k})\gamma_{l}$ and $\sigma_{l}^{x}=i\gamma_{l}\tilde{\gamma}_{l}$, in which the Majorana fermion operators satisfy $\{\gamma_{k},\gamma_{l}\}=\{\tilde{\gamma}_{k},\tilde{\gamma}_{l}\}=2\delta_{kl}$, and $\{\gamma_{k},\tilde{\gamma}_{l}\}=0$. In the Majorana representation,
\begin{equation}
H=-\sum_{l=1}^{L-1}i\tilde{\gamma}_{l}\gamma_{l+1}-h\sum_{l=1}^{L-\alpha}i\tilde{\gamma}_{l}\gamma_{l+\alpha},
\label{eq:Majorana}
\end{equation}
in which $\alpha=0$ for the TFI and $2$ for the CI chain. At the quantum critical point (QCP), both models are mapped to 1D massless Majorana fermions.

In order to diagnolize the Hamiltonian (\ref{eq:Majorana}), we first rewrite it with complex fermion operators $c_{l}$ and $c_{l}^{\dagger}$: $\gamma_{l}=c_{l}^{\dagger}+c_{l}$ and $\tilde{\gamma}_{l}=i(c_{l}^{\dagger}-c_{l})$,
\begin{equation}
H=\sum_{i,j}\Big(c_{i}^{\dagger} A_{ij} c_{j}+\frac{1}{2}\big(c_{i}^{\dagger} B_{ij} c_{j}^{\dagger}+\mathrm{H.c.}\big)\Big),
\end{equation}
in which the matrix $A$ is Hermitian, and $B$ is antisymmetric. We then perform the canonical transformation,
\begin{align}
\eta_{k} &= \sum_{i}(g_{ki}c_{i}+h_{ki}c_{i}^{\dagger}), \\
\eta_{k}^{\dagger} &= \sum_{i}(g_{ki}c_{i}^{\dagger}+h_{ki}c_{i}),
\end{align}
in which $g_{ki}$ and $h_{ki}$ satisfy the orthogonality conditions,
\begin{align}
\sum_{i}(g_{ki}g_{li}+h_{ki}h_{li}) &= \delta_{kl}, \\
\sum_{i}(g_{ki}h_{li}+h_{ki}g_{li}) &= 0,
\end{align}
and the secular equations,
\begin{align}
\sum_{i}(g_{ki} A_{il}-h_{ki} B_{il}) &= \Lambda_{k} g_{kl}, \\
\sum_{i}(g_{ki} B_{il}-h_{ki} A_{il}) &= \Lambda_{k} h_{kl}.
\end{align}
The Hamiltonian is diagonalized,
\begin{equation}
H=\sum_{k} \Lambda_{k} \eta_{k}^{\dagger} \eta_{k},
\end{equation}
in which the eigenstate energy $\Lambda_k\geq 0$, thus the ground state corresponds to the vacuum state of the complex fermions $\eta_k$'s.

The equal-time spin correlation function at the ground state is given by
\begin{equation}
\langle \sigma_{l}^{z}\sigma_{m}^{z}\rangle	=\bigg\langle \prod_{i=l}^{m-1}(i \tilde{\gamma_i} \gamma_{i+1} )\bigg\rangle,\quad l<m.
\end{equation}
According to the Wick theorem, the correlation function of mutually anticommuting operators $\mcal{A}_{i}$'s is given by
\begin{equation}
\langle\mcal{A}_{1}\cdots\mcal{A}_{2n}\rangle=\sum_{\text{all pairings}}(-1)^{\sigma}\prod_{\text{all pairs }(ij)}\langle \mcal{A}_{i}\mcal{A}_{j}\rangle,
\end{equation}
in which $(-1)^{\sigma}$ is the parity of the permutation. Define $\phi_{ki}=g_{ki}+h_{ki}$, $\psi_{ki}=g_{ki}-h_{ki}$, then the pair correlations of Majorana operators are given by
\begin{align}
\langle \gamma_{i} \gamma_{j}\rangle &= \sum_{k} \phi_{ki} \phi_{kj}=\delta_{ij}, \\
\langle \tilde{\gamma}_{i} \tilde{\gamma}_{j}\rangle &= \sum_{k} \psi_{ki} \psi_{kj}=\delta_{ij}, \\
\langle i \tilde{\gamma}_{i} \gamma_{j}\rangle &= \sum_{k} \psi_{ki} \phi_{kj} \equiv G_{ij}.
\end{align}
Therefore, we find
\begin{equation}
\langle\sigma_{l}^{z}\sigma_{m}^{z}\rangle=
\left|
\begin{array}{cccc}
G_{l, l+1} 	& G_{l, l+2} 	& \cdots 	& G_{l,m} 	\\
G_{l+1,l+1} & G_{l+1,l+2}	& \cdots 	& G_{l+1,m} \\
\vdots 		& \vdots		& \ddots	& \vdots 	\\
G_{m-1,l+1}	& G_{m-1,l+2}	& \cdots	& G_{m-1,m}
\end{array}
\right|.
\label{eq:det}
\end{equation}

For the TFI chain, the connected correlation function of the spin operator $\sigma_{l}^{z}$ is given by
\begin{equation}
C_{\perp}(1,1+L/2) = \langle\sigma_{1}^{z}\sigma_{1+L/2}^{z}\rangle.
\end{equation}

For the CI chain, the ground state is two-fold degenerate due to the spontaneous magnetization on the edges. We take the expectation value over an eigenstate of $\sigma_{1}^{z}$, then $\langle\sigma_{1}^{z}\sigma_{l}^{z}\rangle=\langle\sigma_{1}^{z}\rangle\langle\sigma_{l}^{z}\rangle$, thus the connected correlation $C_{\perp}(1,r)=\langle\sigma_{1}^{z}\sigma_{l}^{z}\rangle_{c}=0$. Therefore, we calculate the connected correlation function of the second site with the bulk instead,
\begin{equation}
C_{\perp}(2,2+L/2) = \langle \sigma_{2}^{z} \sigma_{2+L/2}^{z}\rangle-\langle \sigma_{2}^{z}\rangle\langle \sigma_{2+L/2}^{z}\rangle =\langle \sigma_{2}^{z} \sigma_{2+L/2}^{z}\rangle-\langle \sigma_{1}^{z}\sigma_{2}^{z}\rangle\langle\sigma_{1}^{z}\sigma_{2+L/2}^{z}\rangle.
\end{equation}

The connected correlation function of the energy operator $\epsilon_{l}=\sigma_{l}^{z}\sigma_{l+1}^{z}$ is also calculated in the Majorana representation,
\begin{equation}
C_{\perp}(L/2) = \langle \epsilon_{1}\epsilon_{1+L/2}\rangle -\langle\epsilon_{1}\rangle \langle\epsilon_{1+L/2}\rangle = \langle i \tilde{\gamma_1} \gamma_{2+L/2}\rangle\langle	i \gamma_{2}\tilde{\gamma}_{1+L/2}\rangle.
\end{equation}

\begin{figure*}[tb]
\includegraphics[width=0.8\textwidth]{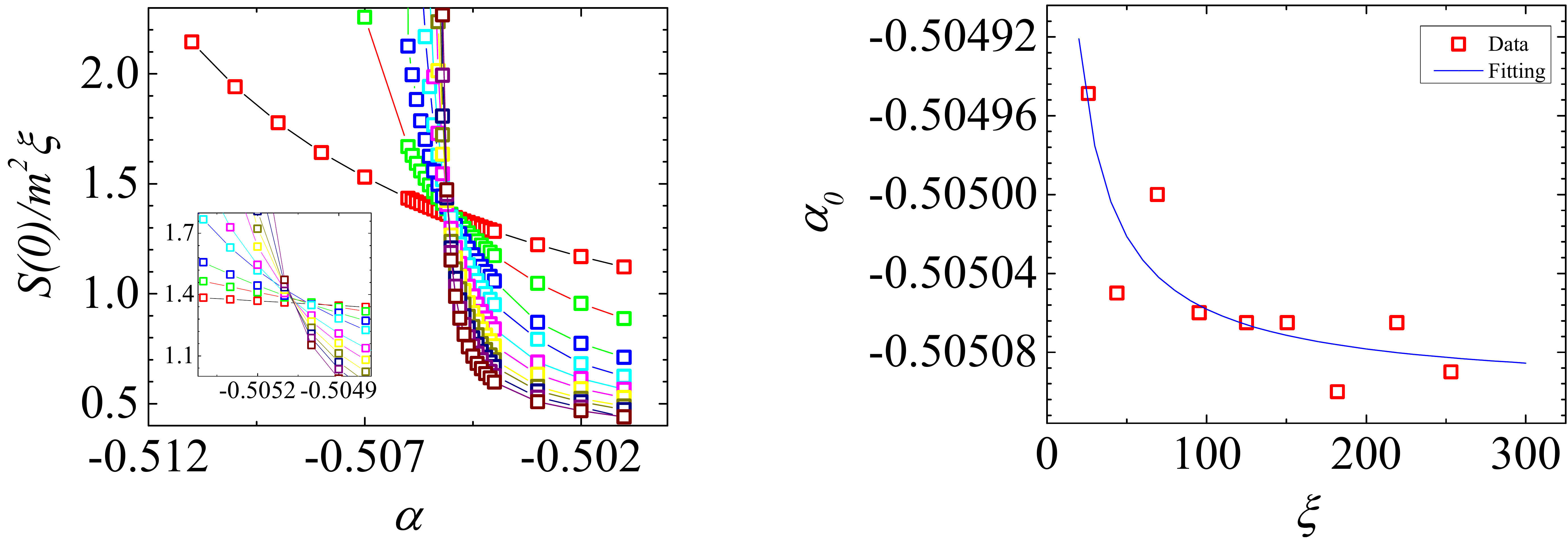}
\caption{The crossing points of the dimensionless quantity $\frac{S(0)}{m^2 \xi}$ calculated with different bond dimension $D$ (left panel) provide estimate of the critical point $\alpha_{c}$, which is extrapolated to the infinite-$\xi$ limit with $\alpha_{c}(\xi)=\alpha_{c}+b\xi^{-\omega}$ (right panel).}
\label{fig_gc}
\end{figure*}

\section{Numerical simulation of the generalized Potts chain}

We apply the matrix product state (MPS) based numerical methods \cite{Verstraete2008, Schollwock2011} to calculate the quantum critical behavior of the generalized Potts chain. The MPS is a class of wavefunctions constructed as the product of a set of $D\times D$ matrices at each site, in which $D$ is called the bond dimension. They represent low-entanglement quantum many-body states faithfully and are the underlying wavefunctions of the density matrix renormalization group (DMRG) algorithm \cite{White1992}. MPS can accurately describe the low-energy properties of one-dimensional (1D) gapped systems \cite{Schollwock2005, Verstraete2006}, while for critical systems, MPS can be adopted for finite-size scaling analysis.

\subsection{Bulk quantum critical point}

We use the variational uniform infinite MPS \cite{Haegeman2017, Zauner-Stauber2018} to determine the RSPT-FM QCP of the generalized three-state Potts chain. The uniform infinite MPS is translationally invariant and defined directly in the thermodynamic limit. For a given uniform infinite MPS with a bond dimension $D$, the correlation length $\xi$ is defined by
\begin{equation}
\xi^{-1}=\ln(\lambda_{0}/\lambda_{1}),
\end{equation}
in which $\lambda_{0}$ and $\lambda_{1}$ are the largest two eigenvalues of the transfer matrix constructed with the local MPS tensor \cite{Haegeman2017, Zauner-Stauber2018}.

We fix $\beta=1$, and calculate the approximate ground states for different $\alpha$ with various bond dimensions $D$. The largest $D$ used in the simulation is $300$. Several random initial states are used to avoid local minima. The structure factor of the order parameter is defined by
\begin{equation}
S(q) = \sum_i e^{iqr_i} \langle \sigma_0 \sigma_i \rangle_c,
\end{equation}
in which the infinite sum can be efficiently calculated with the pseudo-inverse of the transfer matrix \cite{Haegeman2017, Zauner-Stauber2018}. Near the RSPT-FM QCP, $S(0)$ has the same scaling dimension as $m^2 \xi$, in which $m$ is the order parameter. The following dimensionless ratio has the scaling form \cite{Rader2018, Corboz2018},
\begin{equation}
\frac{S(0)}{m^2 \xi} = f(g \xi^{1/\nu}),
\end{equation}
in which $g = \alpha-\alpha_{c}$. As shown in Fig. \ref{fig_gc}, the crossing point of this ratio calculated with different $D$ give the estimate of the critical point $\alpha_{c}$. By extrapolating to the infinite-$\xi$ limit, we find $\alpha_{c} = -0.50509(4)$ for $\beta=1$. The disorder-NotA QCP is obtained by the self-duality mapping as $(\alpha_{c},-1)$.

\subsection{Boundary correlation functions}

\begin{figure*}[tb]
\includegraphics[width=0.8\textwidth]{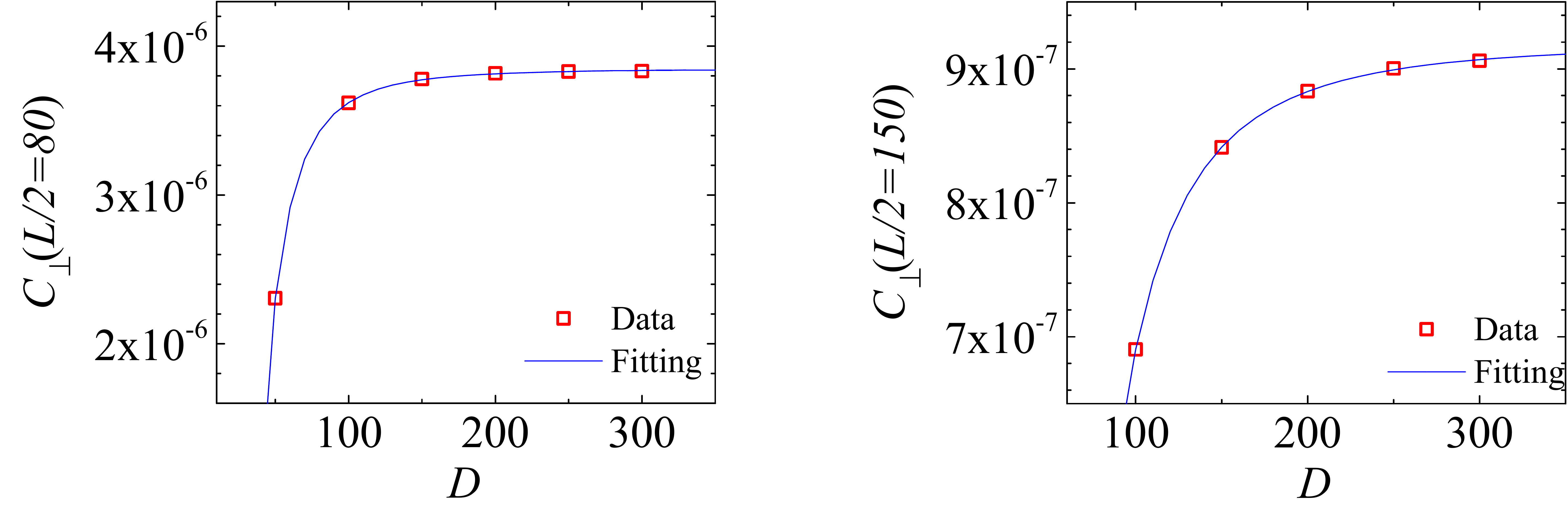}
\caption{Examples of finite-$D$ scaling of the energy correlation function $C_{\perp}(L/2)$ at the QCP of the generalized Potts chain. The data converge to the infinite-$D$ limit exponentially, and are fitted with $c+ae^{-bD}$.}
\label{fig_finite_D}
\end{figure*}

We use the DMRG algorithm \cite{White1992} on finite lattices to calculate the correlation functions of the generalized Potts chain with open boundary condition at the QCPs. For various bond dimensions $D$, we carry out two-site DMRG calculations followed by single-site DMRG sweeps until convergence to achieve the accurate ground state wavefunctions. The correlation functions are calculated for these MPS wavefunctions, and then extrapolated to the infinite-$D$ limit (see Fig. \ref{fig_finite_D}). These extrapolated correlations are then fed into the finite-size scaling analysis in the main text.

\subsection{Entanglement spectra}

\begin{figure*}[tb]
\includegraphics[width=\textwidth]{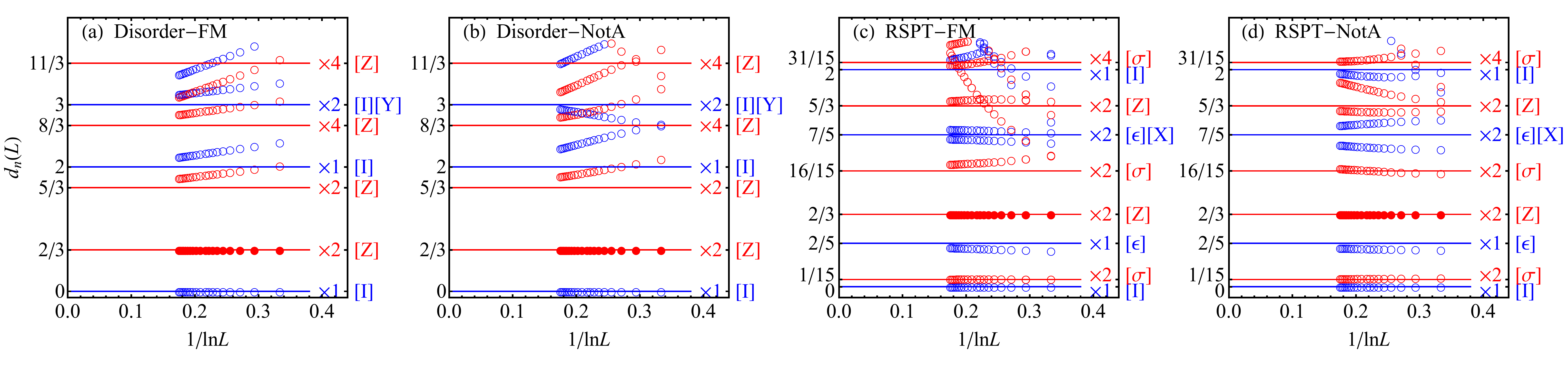}
\caption{Entanglement spectra of the generalized Potts chain at the QCPs. The entanglement gap $\Delta\epsilon_{n}(L)=\epsilon_{n}(L)-\epsilon_{0}(L)$ is normalized to the effective scaling dimension, $d_{n}(L)=\frac{2}{3}\frac{\Delta\epsilon_{n}(L)}{\Delta\epsilon_{Z}(L)}$, such that the primary state in $[Z]$ (marked with filled circles) is normalized to $2/3$. The conformal family and the expected degeneracy in the boundary CFT are labeled on the right. Red circles indicate numerically exactly two-fold degenerate levels, while blue circles indicate non-degenerate levels. The disorder-FM and the disorder-NotA QCPs are consistent with the operator content of the Potts CFT with free b.c. \cite{Cardy1986a}, $[\mathbb{I}]\oplus [Y]\oplus 2[Z]$, while the RSPT-FM and the RSPT-NotA QCPs are consistent with that of the dual-mixed b.c. \cite{Affleck1998, Fuchs1998}, $[\mathbb{I}]\oplus [Y]\oplus[\epsilon]\oplus [X]\oplus 2[\sigma]\oplus 2[Z]$.}
\label{fig:PottsSpecSupp}
\end{figure*}

The operator content of a boundary CFT also shows up in the bipartite entanglement spectrum of the ground state \cite{Lauchli2013, Ohmori2015, Cardy2016}. The entanglement spectrum is equivalent to the energy spectrum of the CFT with a proper conformal b.c. specified at the entangling surface, i.e., the boundary between the subsystem and the rest of the chain \cite{Cardy2016}. It was found in a few 1D quantum critical states that the conformal b.c. at the entangling surface is given by the free b.c. \cite{Lauchli2013} unless a boundary state is explicitly specified by inserting a projection operator in the reduced density matrix \cite{Ohmori2015}.

We calculate the entanglement spectra of the ground state at the four QCPs (Fig. \ref{fig:PottsSpecSupp}). The entanglement spectra at the disorder-FM and the disorder-NotA QCPs are given by the operator content of the Potts CFT with free b.c., which confirms the observation in Ref. \onlinecite{Lauchli2013}. However, we find that the entanglement spectra at the RSPT-FM and the RSPT-NotA QCPs are drastically different and consistent with the Potts CFT with dual-mixed b.c. instead. Therefore, the distinct conformal b.c. is realized at both the physical and the entangling surfaces.

The Affleck-Ludwig boundary entropy is a universal constant in the thermal and the entanglement entropy, which is determined by the energy and the entanglement spectrum of the conformal b.c. \cite{Affleck1991}. The $n$-th R\'enyi entropy of a subsystem of length $\ell$ scales as \cite{Calabrese2004, Calabrese2009b, Cardy2016, Alba2018}
\begin{equation}
S_{n}(\ell)=\frac{c}{12}(1+1/n)\ln(\ell/\epsilon)+\ln g_{A}+\ln g_{B}+O(1/\ell),
\end{equation}
in which $c$ is the central charge, $\epsilon$ is a non-universal short-range cutoff, and $\ln g_{A}$ and $\ln g_{B}$ are the universal boundary entropy of the two boundaries. If there are degenerate edge states, they contribute an integer factor of degeneracy to $g_{A}g_{B}$ \cite{Alba2018} similar to the topological entanglement entropy of gapped states \cite{Kitaev2006, Levin2006}. For generic conformal b.c., $g_{A}g_{B}$ is not an integer but may be taken as a ``generalized degeneracy''. Therefore, the conformal b.c. generalizes the edge degeneracy in characterizing the quantum critical states.

\bibliography{../../Bibtex/library}